\documentclass[prd,showpacs,preprintnumbers,amsmath,amssymb,floatfix]{revtex4}

\usepackage{graphicx}
\usepackage{dcolumn}
\usepackage{bm}
\usepackage{amssymb}
\usepackage{epsfig}
\usepackage{color}

\newcommand{\ba}{\begin{eqnarray}}
\newcommand{\ea}{\end{eqnarray}}
\newcommand{\be}{\begin{equation}}
\newcommand{\ee}{\end{equation}}
\newcommand{\bdisplay}{\begin{displaymath}}
\newcommand{\edisplay}{\end{displaymath}}

\newcommand{\eq}[1]{Eq.\,(\ref{#1})}
\newcommand{\fig}[1]{Fig.\,\ref{#1}}

\begin{document}

\title{Comment on ``More on Heisenberg's model for high energy nucleon-nucleon scattering'' }

\author{Martin~M.~Block}
\email{mblock@northwestern.edu}
\affiliation{Department of Physics and Astronomy, Northwestern University,
Evanston, IL 60208}
\author{Loyal Durand}
\email{ldurand@hep.wisc.edu}
\altaffiliation{Present address: 415 Pearl Ct., Aspen, CO 81611}
\affiliation{Department of Physics, University of Wisconsin, Madison, WI 53706}
\author{Phuoc Ha}
\email{pdha@towson.edu}
\affiliation{Department of Physics, Astronomy and Geosciences, Towson University, Towson, MD 21252}
\author{Francis Halzen}
\email{francis.halzen@icecube.wisc.edu}
\affiliation{Wisconsin IceCube Particle Astrophysics Center and Department
of Physics, University of Wisconsin-Madison,  Madison,  Wisconsin 53706}

\begin{abstract}

We comment on the treatment of asymptotic black-disk scattering in a recent paper of Nastase and Sonnenschein, Phys.\ Rev.\ D\ {\bf 92}, 015028 (2015),   on scattering in an updated version of the Heisenberg model which gives $pp$ and $\bar{p}p$ cross sections which increase at very high energies as $\ln^2s$. We show that the total cross section they define does not correspond to that measured in experiments, with the result that their limit for the ratio $\sigma_{\rm elas}/\sigma_{\rm tot}$ is too small by a factor 2. The correct ratio for black-disk scattering, $\sigma_{\rm elas}/\sigma_{\rm tot} \rightarrow 1/2$ for $s\rightarrow\infty$, is strongly supported by experiment.

\end{abstract}


\maketitle

%%%%%%%%%%%%%%%%%%%%%%%%%
%%%%%%%%%%%%%%%%%%%%%%%%%

In an interesting recent paper on Heisenberg's model for nucleon-nucleon scattering which gives a total cross section that increases as $\ln^2s$ at very high energies, Nastase and Sonnenschein \cite{nastase} studied the possible approach to ``black-disk'' scattering.  Their result, that $\sigma_{\rm elas}/\sigma_{\rm tot}\rightarrow 1/4$ for $s\rightarrow \infty$ in this model, which they supported by reference to the recent TOTEM data \cite{totem2011,totem2013_2} from the Large Hadron Collider (LHC), disagrees with the standard black-disk result,  $\sigma_{\rm elas}/\sigma_{\rm tot}\rightarrow 1/2$. The latter is strongly favored by a comprehensive fit \cite{bdhhfit} to all the $pp$ and $\bar{p}p$  data including those on total, elastic, and inelastic cross sections,  the ratio $\rho$ of the real to the imaginary parts of the forward scattering amplitudes, and on the forward slope parameters $B$ for the differential elastic scattering cross sections.  We show here that the disagreement arises from an incorrect definition of the total cross section that does not correspond to the cross section actually measured. We also summarize the experimental  evidence for a black-disk limit in $pp$ and $\bar{p}p$ scattering.

The standard relation between the total scattering cross section and the imaginary part of the forward elastic scattering amplitude $f(s,t)$ given by the optical theorem  follows from the unitarity of the $S$ matrix for the scattering \cite{blockcahn}, $S^\dagger S=\openone$, or writing $S$ as $S=\openone+iT$ , $i\left(T^\dagger-T\right)=T^\dagger T$. Specializing to the case of forward elastic two-body scattering $|i\rangle\rightarrow|i\rangle$, this gives
\be
\label{optical}
 2\,{\rm Im}\,T_{ii}
= \int(2\pi)^4\delta^4(P_{i'}-P_i)\left|T_{i'i}\right|^2d\rho_{i'} +\sum_{k\not=i'} \int(2\pi)^4\delta^4\left(P_k-P_i\right)T^*_{ki}T_{ki}d\rho_k,
\ee
where the state $|i'\rangle$ differs from the initial state $|i\rangle$ only in the momenta of the scattered particles, and  the sums over the final states include integrations $d\rho_k$ over the phase space of the final-state  $|k\rangle$ subject to the conservation of the total 4-momenta $P$.

The right-hand side of \eq{optical} is the sum of the transition rates from the initial state $|i\rangle$ to all possible final states. Dividing by the incident flux  $4p\sqrt{s}$ in the initial state, we obtain the total  cross section $\sigma_{\rm tot}= {\rm Im}\,T_{ii}/2p\sqrt{s}$, where $p$ and $s$ are the momentum of either particle and the square of the total energy in the center-of-mass system, and $t=-2p^2(1-\cos{\theta})$ is the square of the 4-momentum transfer in the scattering, with $\theta$  the scattering angle. The  elastic scattering cross section is given by the first term in the sum, $\sigma_{\rm elas}=\int d\Omega\, |T_{i'i}|^2/64\pi^2s$. The remainder of the sum gives $\sigma_{\rm inel}$. These definitions correspond to experimental practice, where the cross sections are defined as the ratios of the yields or interactions per unit time to the incident fluxes.

We will write the differential  elastic scattering cross section $d\sigma_{\rm elas}/d\Omega$ in terms of the scattering amplitude  $f(s,t)=T_{i'i}/8\pi\sqrt{s}$  with $d\sigma_{\rm elas}/d\Omega=|f(s,t)|^2$ \footnote{Note that the normalization used for $f(s,t)$ in \cite{bdhhfit} is to $d\sigma_{\rm elas}/dt$ rather than to $d\sigma_{\rm elas}/d\Omega$, hence the factor $p$ in \eq{f_defined} relative to Eq.\ (1) in \cite{bdhh_eikonal}.}. With this definition, \eq{optical} gives the familiar relation $\sigma_{\rm tot}=(4\pi/p)\,{\rm Im}\,f(s,0)$---the optical theorem---for $pp$ or $\bar{p}p$ scattering.

Given angular momentum conservation, the scattering amplitude can be expanded in the usual partial-wave series, $f(s,t)=\sum_j (2j
+1)f_j(s)P_j(\cos{\theta})$. The sum can be converted to an integral at high energies, where many partial waves $j$ contribute to the scattering,  by using the relation $P_j(\cos{\theta})\approx J_0(b\sqrt{-t})$ where  $pb=\sqrt{j(j+1)}$. This gives the impact-parameter or eikonal  representation of the scattering amplitude \cite{blockrev},
\be
\label{f_defined}
f(s,t)=ip\int_0^\infty db\,b\left(1-e^{i\chi(b,s)}\right)J_0\left(b\sqrt{-t}\right).
\ee
Here  $S_{ii}=e^{i\chi(b,s)}$ is written in terms of its magnitude and phase through the complex eikonal function $\chi(b,s)=\chi_R(b,s)+i\chi_I(b,s)$ with $\chi_I\geq 0$. Using the optical theorem, we obtain the total cross section as
\be
\label{sigma_tot}
\sigma_{\rm tot}(s)=4\pi\int_0^\infty db\,b\left(1-\cos{\chi_R(b,s)}e^{-\chi_I(b,s)}\right).
\ee

We obtain the elastic scattering cross section by squaring $f(s,t)$ and integrating over angles using the orthogonality relations for the Bessel functions,
\be
\label{sigma_elas}
\sigma_{\rm elas}(s)=2\pi\int_0^\infty db\,b\left(1-2\cos{\chi_R(b,s)}e^{-\chi_I(b,s)}+e^{-2\chi_I(b,s)}\right).
\ee
The inelastic cross section is given by the difference between \eq{sigma_tot} and \eq{sigma_elas},
\be
\label{sigma_inel}
\sigma_{\rm inel}=2\pi\int_0^\infty db\,b\left(1-e^{-2\chi_I(b,s)}\right).
\ee

In the high-energy limit, we expect the scattering to be strongly absorptive with $e^{-\chi_I(b,s)}\approx 0$ out to a maximum impact parameter $b_{\rm max} =R(s)\propto \ln\left(\sqrt{s}\right)$ \cite{bdhh_eikonal,bdhhfit}, and to be essentially negligible at larger impact parameters. The effect of the real part of the eikonal function through the factors $\cos{\chi_R}$ is small, with $(1-\cos{\chi_R})e^{-\chi_I}\approx 0$ except in a narrow ``edge'' region centered on $R(s)$ \cite{bdhh_eikonal,edge}, with its overall contribution vanishing as $1/R(s)\propto 1/\ln{s}$ relative to the cross sections. Replacing $\cos{\chi_R}$ by 1 and integrating Eqs.\ (\ref{sigma_tot}) and (\ref{sigma_elas}) from $b=0$ to $R(s)$, we obtain the so-called ``black disk'' limit corresponding to scattering from a completely absorbing disk of radius $R(s)$, with
$\sigma_{\rm tot} \rightarrow 2\pi R^2$, $\sigma_{\rm elas}\rightarrow \pi R^2$, and $\sigma_{\rm elas}/\sigma_{\rm tot}\rightarrow 1/2$. Similarly, $\sigma_{\rm inel}\rightarrow \pi R^2$ and $\sigma_{\rm inel}/\sigma_{\rm tot}\rightarrow 1/2$.

The inelastic cross section in this limit, familiar in wave optics for strong absorption and short wavelengths, corresponds to the geometrical area of the target. The elastic cross section arises entirely from wave diffraction around the absorbing region, with the cross sections equal (Babinet's principle), so the total cross section is twice the absorption cross section.

In the case of nucleon-nucleon scattering, the black-disk radius $R(s)$ is expected to grow as $\ln\left(\sqrt{s}\right)$ at high energies, consistently with the Froissart-Martin \cite{froissart,martin1} bound on total cross sections. This behavior is characteristic of models in which $\chi_I(b,s)$ grows in magnitude as a power of $s$ but is cut off exponentially in its $b$ dependence at large $b$ \cite{bdhh_eikonal}.

The Heisenberg model studied in \cite{nastase} gives an example. In that model, the growth of the cross section is ascribed to the growth in pion production with increasing energy.  The maximum impact parameter at which pion production occurs for a cutoff $\propto e^{-m_\pi b}$ determines $R(s)$, giving  $R(s)=\left(\pi/m_\pi^2\right)\ln\left(\sqrt{s}/\langle k_0\rangle\right)$ where $\langle k_0\rangle$, the average energy per pion of the pions emitted in the collision, is assumed by Heisenberg to be constant.

The cross section $\sigma=\pi b_{\rm max}^2=\pi R^2$ calculated in \cite{nastase} following Heisenberg  is that for pion production---the absorptive or inelastic cross section---and not the total cross section. It was unfortunately taken as the total cross section in \cite{nastase}, and the optical theorem then used to determine the normalization of their model amplitude for elastic black-disk scattering.
As a result,  the scattering amplitudes used in the analysis of  black-disk scattering in Sec.\ VIII\,B of that paper are not consistent with the discussion above, with the result that their total cross section is too small by a factor of 2, and their elastic cross section, by a factor of 4. This gives the asymptotic ratio $\sigma_{\rm elas}/\sigma_{\rm tot}\rightarrow 1/4$, and, for $\sigma_{\rm tot}=\sigma_{\rm elas}+\sigma_{\rm inel}$,  $\sigma_{\rm inel}/\sigma_{\rm tot}\rightarrow 3/4$. As they note, these predictions differ from the usual ratios 1/2.

Aside from the theoretical problem, the predicted ratios are not consistent with experiment.
In \cite{bdhhfit}, we have presented strong evidence that the $pp$ and $\bar{p}p$ scattering amplitudes approach those expected for scattering from a completely absorptive or ``black'' disk with a logarithmically increasing radius $R(s)$ at very high energies. As noted above, this conclusion was based on a fit  to the high-energy $pp$ and $\bar{p}p$ scattering  data on total, elastic, and inelastic cross sections,  $\rho$ values, and forward slope parameters $B$ for the differential cross sections. The expressions  used for the cross sections were quadratic in $\ln{s}$, with extra rapidly falling Regge-like terms that are significant only at lower energies. The total cross sections were constrained to match smoothly to the low-energy cross sections at 4 GeV. Details are discussed in \cite{bdhhfit}.

Comparing the coefficients of the terms in $\ln^2s$ in $\sigma_{\rm tot}$, $\sigma_{\rm elas}$, and $B$, we found in a fit unconstrained at high energies that $\sigma_{\rm elas}/\sigma_{\rm tot}\rightarrow 0.528\pm 0.108$ and $B\rightarrow (0.990\pm 0.415)\sigma_{\rm tot}/8\pi$, consistent with the expected results $\sigma_{\rm elas}/\sigma_{\rm tot}\rightarrow 1/2$ and $B\rightarrow \sigma_{\rm tot}/8\pi$. If we imposed these conditions as extra, high-energy, constraints in the fitting procedure, we obtained an equally good fit, with significantly reduced uncertainties.

%%%%%%%%%%%%%%%%%%%%%
%%%%%% FIG 1 cross sections no HE constraints  %%%%%

\begin{figure}[htbp]
\includegraphics{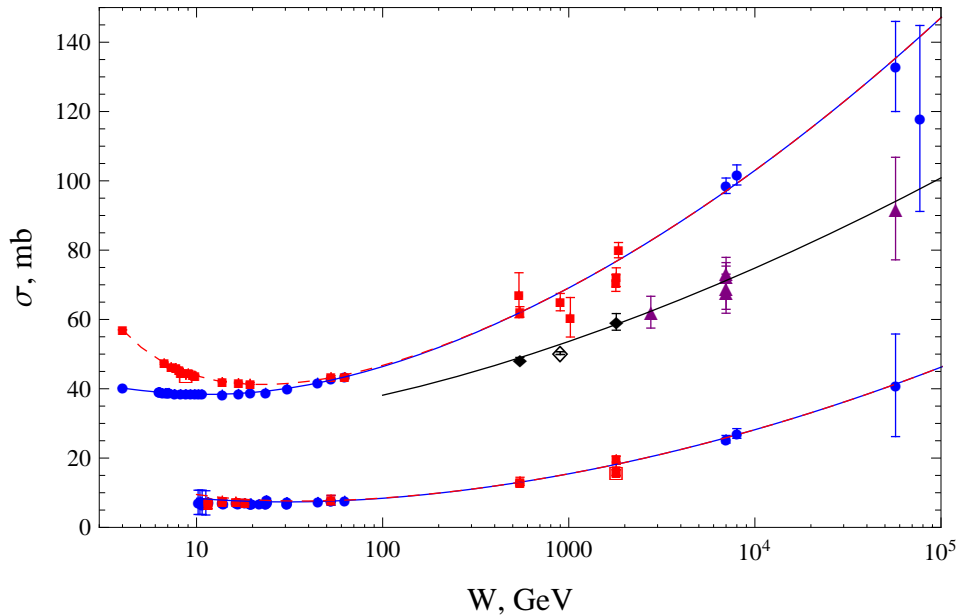}
\caption{Fits, top to bottom, to the total, inelastic, and elastic scattering cross sections  using the low-energy analyticity constraints on the  cross sections discussed in \cite{bdhh_eikonal}, plotted as functions of the center-of-mass energy $W=\sqrt{s}$: $\sigma_{\rm tot}^{\bar{p}p}$ and  $\sigma_{\rm elas}^{\bar{p}p}$ (red) squares and dashed (red) line; $\sigma_{\rm tot}^{pp}$  and $\sigma_{\rm elas}^{pp}$ (blue) dots and solid (blue) line; $\sigma_{\rm inel}^{\bar{p}p}$ (black) diamonds and line; $\sigma_{\rm inel}^{pp} $ (purple) triangles.  The fit used only data  on center-of-mass energies  $W\geq 6$ GeV for $\sigma_{\rm tot}$,  $W\geq 30$ GeV for $\sigma_{\rm elas}$, and $W\geq 540$ GeV for $\sigma_{\rm inel}$.  The curve for $\sigma_{\rm elas}$ includes data down to 10 GeV to show how the cross section is tied down at lower energies.  Outlying points not used in the fit are shown with large open symbols surrounding the central points; the size of those symbols does not reflect the quoted errors of the measurement. }
\label{fig:xsectionsnoBD}
\end{figure}
%%%%%%%%%%%%%%%%%%%%%
%

We show the fit we obtained to the cross sections without using the high-energy constraints in \fig{fig:xsectionsnoBD}. It is clearly quite good, with a $\chi^2$ per degree of freedom of 1.2 for 145 degrees of freedom.

\begin{figure}[htbp]
\includegraphics{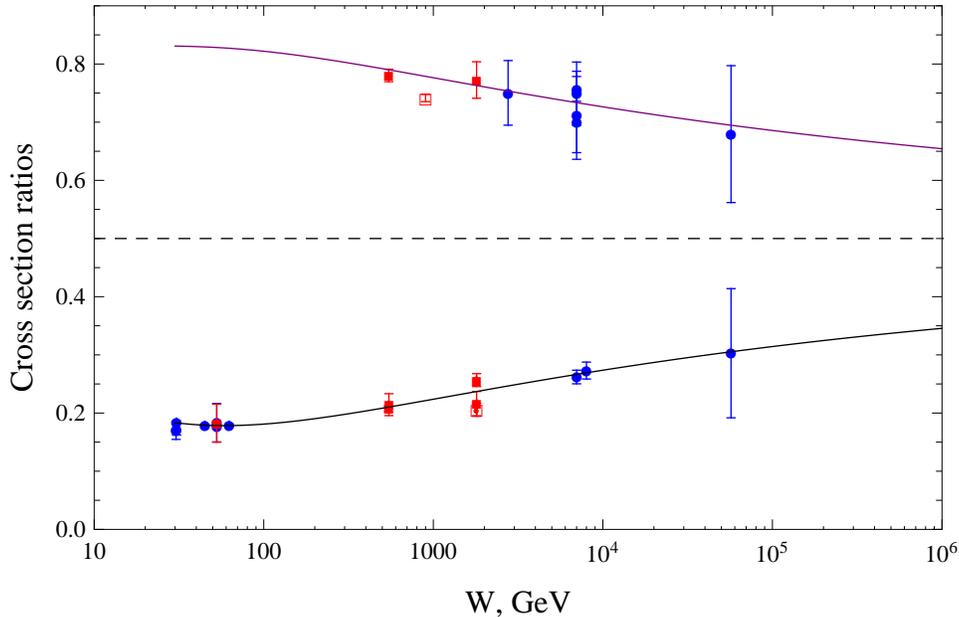}
\caption{The ratios  $\sigma_{\rm inel}/\sigma_{\rm tot}$ (top curves) and $\sigma_{\rm elas}/\sigma_{\rm tot}$ (bottom curves) for $pp$ scattering (blue dots) and $\bar{p}p$ scattering (red squares), plotted as functions of the center-of-mass energy $W=\sqrt{s}$.  Outlying points not used in the fit are shown with large open symbols surrounding the central points as in \fig{fig:xsectionsnoBD}. The plotted ratios use the cross sections with only low-energy constraints from \cite{bdhh_eikonal}. The points shown for  $\sigma_{\rm inel}/\sigma_{\rm tot}$ and $\sigma_{\rm elas}/\sigma_{\rm tot}$ are from  independent measurements of $\sigma_{\rm inel}$ and $\sigma_{\rm elas}$ and the fitted $\sigma_{\rm tot}$.}
\label{fig:sigmaratios}
\end{figure}

In \fig{fig:sigmaratios}, we show the ratios of the independently measured elastic and inelastic cross sections to the fitted $\sigma_{tot}$. The convergence of the fitted ratios $\sigma_{\rm elas}/\sigma_{\rm tot}$ and $\sigma_{\rm inel}/\sigma_{\rm tot}$ toward the common value 1/2 is clearly evident. We note also that  $\sigma_{\rm elas}/\sigma_{\rm tot}$ exceeds 1/4 at Tevatron energies of 1800 GeV and above. We emphasize that these results follow from a comprehensive analysis of all the high-energy data, and do not depend simply on the few highest-energy points. Indeed, the values of the cross sections measured at the Large Hadron Collider at 7 and 8 TeV and in the AUGER cosmic ray experiment at 57 TeV had been predicted successfully using a model of this type in an analysis that used data only up to 1800 GeV \cite{blockhalzen}, all that existed at the time.

We conclude that the data---including the newer points---are inconsistent with the prediction in \cite{nastase}. The problem in those authors' treatment of the total and elastic scattering cross sections and the black-disk limit of nucleon-nucleon scattering does not appear to affect the remainder of that paper.

%%%%%%%%%%%%%%%%%%%%%
%%%%%% FIG 2 sigma ratios  %%%%%

%%%%%%%%%%%%%%%%%%%%%
%

%%%%%%%%%%%%%

\begin{acknowledgments}

M.M.B., L.D., and F.H.\  would  like to thank the Aspen Center for Physics for its hospitality and for its partial support of this work under NSF Grant No. 1066293. F.H.'s research was supported in part by the U.S. National Science
Foundation under Grants No.~OPP-0236449 and PHY-0969061 and by the
University of Wisconsin Research Committee with funds granted by the Wisconsin Alumni Research
Foundation.   P.H.\ would like to thank Towson University Fisher College of Science and Mathematics for support.

\end{acknowledgments}

\bibliography{small_x_references}

\end{document}